\documentclass[12pt]{article}

\usepackage{amssymb}

\begin{document}

\title{ The Killing scalar of non-linear $\sigma$ model on G/H realizing the 
classical exchange algebra}

\author{Shogo Aoyama\thanks{e-mail: spsaoya@ipc.shizuoka.ac.jp}  \\
       Department of Physics \\
              Shizuoka University \\
                Ohya 836, Shizuoka  \\
                 Japan}
                 
\maketitle
                 
\begin{abstract}
The Poisson brackets for non-linear $\sigma$ models on G/H are set up on the light-like plane. A quantity which transforms irreducibly by the Killing vectors, called the Killing scalar, is constructed  in an arbitrary representation of G.  It is shown to satisfy the classical exchange algebra. 

\end{abstract}

\newpage

\section {Introduction}
\setcounter{equation}{0}

The Yang Baxter equation arises in a large class of exactly solvable models, such as lattice models, spin-chain system, nonlinear $\sigma$-models, conformal field theory, etc. Among them the Yang Baxter equation for the PSU($2,2|4$) spin-chain system  gained much interest in the last decade. Finding its solution lead a discovery of string/QCD duality, namely a relationship  between the PSU($2,2|4$) spin-chain system and the $N=4$ supersymmetric QCD\cite{Bei1,Bei2}. 

To explain the Yang Baxter equation and the resulting  algebraic structure, we take a generic spin-system equipped  with a set quantum operators, say,  $\Psi$.  We consider a tensor product chain of  $\Psi$s and exchange two  of them in 
an adjacent position, say,  $\Psi(x)$ and $\Psi(y)$. Then there may exists a R-matrix  defining  a quantum exchange algebra 
\begin{eqnarray}
R_{xy}\Psi(x)\otimes \Psi(y)=\Psi(y)\otimes \Psi(x).   \label{QEA}
\end{eqnarray} 
 such that it satisfies the Yang-Baxter equation
\begin{eqnarray}
 R_{xy}R_{xz}R_{yz}=R_{yz}R_{xz}R_{xy}
\label{YBB}
\end{eqnarray}
Suppose the  R-matrix to be a quantum deformation of a certain classical r-matrix
as
$$
R_{xy}= 1+ hr_{xy}+O(h^2),
$$
with an infinitesimal parameter $h$. 
Then (\ref{YBB}) becomes  a classical exchange algebra
\begin{eqnarray}
 \{\Psi(x)\mathop{,}^\otimes\Psi(y)\}= -hr_{xy}\Psi(x)\otimes \Psi(y), 
\label{CEA0}
\end{eqnarray}
and the classical Yang-Baxter equation
\begin{eqnarray}
 [r_{xy},r_{xz}]+[r_{xy},r_{yz}]+[r_{xz},r_{yz}]=0.   \label{YB}
\end{eqnarray}
The quantity in the l.h.s. of (\ref{CEA0}) is a Poisson bracket which was substituted for the commutator $\displaystyle{[\Psi(x)\mathop{,}^\otimes\Psi(y)]}$. 
In the recent works\cite{Ao1,Ao2} the classical exchange algebra (\ref{CEA0})  was shown for a classical quantity, called  G-primary, in the constrained WZWN model on the coset space G/\{S$\otimes$U$(1)^d$\}. Namely the G-primary was  constructed out of basic fields of the model in an arbitrary representation of G. By setting up the Poisson brackets for the basic fields, the classical exchange algebra (\ref{CEA0}) was shown to appear with the r-matrix in an arbitrarily chosen representation of the G-primary\cite{Ao2}. It would be promoted to 
the quantum exchange algebra (\ref{YBB}) by the usual quantization of the constrained WZWN model. 
Or in mathematics  the algebraic construction of the R-matrix is known when 
the R-matrix  exists in the the Hopf algebra $A$ in such a way that $R\in A\otimes A$ and 
\begin{eqnarray}
&i. & \quad \Delta'(a)=R\Delta(a) R^{-1}, \quad \forall a\in A,   \nonumber \\
&ii. & \quad (\Delta\otimes 1)(R)=R_{xz}R_{yz},\quad\quad 
(1\otimes \Delta)(R)=R_{xz}R_{xy}.   \nonumber
\end{eqnarray}
Here $\Delta$ is the coproduct and $\Delta'=P\circ\Delta$ with the permutation map $P$. The Yang-Baxter equation is derived as one of the properties of this Hopf algebra\cite{Pres}.

In this letter we  discuss the exchange  algebra 
 for the ordinary non-linear $\sigma$-model on G/H as well.  Generalizing the arguments for the constrained WZWN model on G/\{S$\otimes$U$(1)^d$\}\cite{Ao1,Ao2} to this case is not straightforward. 
  To see a difference  between the two cases , we look  a bit more closely  the arguments done in \cite{Ao1,Ao2}. The fundamental Poisson bracket of the WZWN model was set up for the basic field 
\begin{eqnarray}
 g=g_{\rm L}g_{\rm H}.   \label{gauged}
\end{eqnarray}
This basic field  was  obtained  by reparametrizing  the field of the  WZWN model as $g=g_{\rm L}g_{\rm H}g_{\rm R} \in$ G in accordance with the Gau\ss\  decomposition of the Lie-algebra of G under H$\equiv$S$\otimes$U$(1)^d$ 
$$
\{T_{\rm G/H},T_{\rm H}\}=\{T_{\rm L},T_{\rm R},T_{\rm H}\},
$$
 and constraining $g_{\rm R}$ to be $1$. 
The G-primary which has the conformal weight $0$ and linearly  transforms  in an arbitrary representation of G by the Killing vectors  was constructed from the basic field (\ref{gauged}). It satisfied the classical exchange algebra (\ref{CEA0}) when calculated the Poisson bracket by means of the fundamental one  for (\ref{gauged}). 
We may wonder if such a classical exchange algebra exists for 
the ordinary non-linear $\sigma$-model on G/H, which  has a different reparametrization from (\ref{gauged}). Namely the basic field of the non-linear $\sigma$-model is reparametrized as 
\begin{eqnarray}
g=g_{\rm G/H},    \label{para}
\end{eqnarray}
which is an element $g\in$ G generated by $T_{\rm L}$ and $T_{\rm R}$. In this letter we will show that  the classical exchange algebra (\ref{CEA0}) exists also  for this case.  Indeed we will find a quantity satisfying the classical exchange algebra in an arbitrary representation of G, such as the G-primary for the constrained  WZWN model. But the quantity is called Killing scalar this time, because the ordinary non-linear $\sigma$-model has no longer conformal symmetry and the construction differs from that of the G-primary.

The reader might think about studying the issue in the same way as for the G-primary in the constrained WZWN model, i.e., by  gauge-fixing a gauged WZWN model so that the basic field gets  reparametrized as $g=g_{\rm L}g_{\rm R}$ instead of (\ref{gauged}). For that case the gauged  action has vector-like invariance. It 
no longer has   the chiral invariance, which  played an essential role  for the classical exchange algebra of the G-primary. (See (10) in \cite{Ao2} and (3.1) in \cite{Wit} for the respective gauged action.) 
 The strategy to study the exchange algebra for the G-primary in \cite{Ao1,Ao2} does not work for the case with vector-like gauge invariance. Hence we need to develop a new strategy  for the ordinary non-linear $\sigma$-model on G/H. 

 So it is a central issue of this letter to construct the Killing scalar for the ordinary non-linear $\sigma$-model, which plays the same role  as the G-primary for the constrained WZWN model. It is discussed in section 2. There we find that the Wilson line operator is necessary for the construction in a generic representation of G.  It was not needed for the construction of the G primary at all. In section 3 we discuss the Poisson structure of the ordinary non-linear $\sigma$-model on the light-like plane. The fundamental Poisson brackets are set up consistently on that plane.  By means of them we show the classical exchange algebra for the Killing scalar.

\section{The Killing scalar}
\setcounter{equation}{0}

 We begin by giving a general account  on the ordinary non-linear $\sigma$-model to an  extent such that it is needed for arguments for this letter.
 The standard reparametrization of the coset space G/H is given by the CCWZ formalism\cite{CCWZ} in the following procedure.\footnote {The BKMU formalism is also useful formalism when the coset space admits the complex structure. It is straightforward to adapt the arguments in this section to  the BKMU formalism\cite{BKMU}.}
 Decompose the generators of G as
\begin{eqnarray}
\{T^A\} =\{X^i,H^I\},  \label{T}
\end{eqnarray}
in which $H^I$ are generators of the homogeneous group H, while $X^i$ broken ones. Then  we consider a unitary quantity as the basic field (\ref{gauged})
\begin{eqnarray}
U= e^{i\phi^1 X^1+i\phi^2 X^2+\cdots }\equiv e^{i\phi\cdot X}.  \label{U}
\end{eqnarray}
Here $\phi^1,\phi^2,\cdots $  were introduced  correspondingly  to the broken generators $X^i$ and they are real coordinates reparametrizing the coset space G/H, denoted by $\phi^a$. The Cartan-Maurer $1$-form $U^{-1}dU $ is valued in the Lie-algebra of G as 
\begin{eqnarray}
U^{-1}dU =(e^i_a  X^i + \omega^I_a H^I)d\phi^a. \label{Cartan}
\end{eqnarray}
This defines the vielebeine $e^i_a$   and the connection $\omega^I$ in the local frame of the coset space G/H. 
For an element $e^{i\epsilon^AT^A}$ with real parameters $\epsilon^A$ there exists a compensator $e^{i\rho(\phi,\epsilon)}\in$ H such that 
\begin{eqnarray}
e^{i\epsilon^AT^A}U(\phi)=U(\phi')e^{i\rho(\phi,\epsilon)}.  \label{transf}
\end{eqnarray}
This defines a transformation of the coordinates $\phi^a\rightarrow {\phi'}^a(\phi)$.
 When $\epsilon^A$ are infinitesimally small, this relation defines the Killing vectors $R^{Aa}$ as 
\begin{eqnarray}
\delta \phi^a={\phi'}^a(\phi)-\phi^a\equiv \epsilon^AR^{Aa}(\phi).     \label{Killing}
\end{eqnarray}
The Jacobi-identity for the transformation gives the Lie-algebra of G
\begin{eqnarray}
R^{Aa}R^{Bb}_{\ \  ,a}-R^{Ba}R^{Ab}_{\ \ ,a} =f^{ABC}R^{Cb}, \label{Lie-algebra}
\end{eqnarray}
which can be written as 
$
{\cal L}_{R^A}R^{Bb}= f^{ABC}R^{Cb} 
$
by using the Lie-variation. 

So far our arguments are irrelevant to the representation of G. Let us choose it to be  an $N$-dimensional irreducible representation of G. Under the subgroup H let it to be decomposed into irreducible ones  as 
\begin{eqnarray}
\mbox{\boldmath $N$}= \mbox{\boldmath $N$}^{w_1}\oplus \mbox{\boldmath $N$}^{w_2}\oplus\cdots\oplus\mbox{\boldmath $N$}^{w_{m-1}}\oplus\mbox{\boldmath $N$}^{w_m}.   
\label{decomp}
\end{eqnarray}
Here $\mbox{\boldmath $N$}^{w_\mu}, \mu=1,2,\cdots,m$, denote the $N_\mu$-dimensional representation of H with  a set of weight vectors $w_\mu$. 
 Accordingly  $U$ is represented  by an $N\times N$ matrix ${\cal D}(U)$ in a decomposed form  as 
\begin{eqnarray}
{\cal D}(U)=\left(
\begin{array}{cccc}
\noalign{\vskip0.2cm}
(U)_{\scriptscriptstyle{N_1\times N_1}}  & (U)_{\scriptscriptstyle{N_1\times N_2}}  & \cdots   & (U)_{\scriptscriptstyle{N_1\times N_m}} \\
\noalign{\vskip0.2cm} 
(U)_{\scriptscriptstyle{N_2\times N_1}} &(U)_{\scriptscriptstyle{N_2\times N_2}} & \cdots  &  (U)_{\scriptscriptstyle{N_2\times N_m}}      \\
\noalign{\vskip0.2cm}
\vdots   &\hspace{-0.2cm}  \vdots   &  \ddots  &\hspace{-0.2cm} \vdots    \\
\noalign{\vskip0.2cm}
 (U)_{\scriptscriptstyle{N_m\times N_1}}     &\hspace{-0.2cm} (U)_{\scriptscriptstyle{N_m\times N_2}}    &  \cdots    &  (U)_{\scriptscriptstyle{N_{m}\times N_{m}} }    \\
\noalign{\vskip0.2cm}
\end{array}
\right),    \label{D(U)}
\end{eqnarray}
in which $(U)_{N_\mu\times N_\nu}$ is  an $N_\mu\times N_\nu$ matrix. 
From this we take out a set of column vectors such as 
\begin{eqnarray}
\Psi =\left(
\begin{array}{c}
\noalign{\vskip0.2cm}
(U)_{\scriptscriptstyle{N_1\times N_\nu}}   \\
\noalign{\vskip0.2cm} 
(U)_{\scriptscriptstyle{N_2\times N_\nu}}   \\
\noalign{\vskip0.2cm}
 \vdots    \\
\noalign{\vskip0.2cm}
 (U)_{\scriptscriptstyle{N_m\times N_\nu}}     \\
\noalign{\vskip0.2cm}
\end{array}
\right).    \label{Psi}
\end{eqnarray}
Then it transform as $(\mbox{\boldmath $N$},\mbox{\boldmath $N$}^{w_\nu})$ by the  transformation defined by (\ref{transf}), i.e.,
\begin{eqnarray}
\Psi\longrightarrow (e^{i\epsilon^AT^A})_{N\times N}\Psi 
(e^{-i\rho(\phi,\epsilon)})_{N_\nu\times N_\nu},\quad\quad \nu=1,2,\cdots,m.
 \label{Psitransf}
\end{eqnarray}
 If $\mbox{\boldmath $N$}^{w_\nu}$ happens to be a singlet, then $\Psi$ transforms as $\mbox{\boldmath $N$}$, 
i.e.,
\begin{eqnarray}
\delta \Psi = \epsilon^A {\cal D}(T^A)\Psi. \label{KillingS}
\end{eqnarray}
We call this quantity  Killing scalar since (\ref{KillingS}) can be written as  $$
{\cal L}_{R^A}\Psi={\cal D}(T^A)\Psi
$$ 
by using the Lie-variation. For the the coset space SO(6)/SO(5)(=S$^5$) 
 the Killing scalar transforming as the fundamental representation ${\bf 6}$ of SO(6) was given in \cite{Ao3}. The decomposition of ${\bf 6}$ under SO(5) indeed contains a singlet. 
But a generic representation $\mbox{\boldmath $N$}$ of G does not contain it 
in the decomposition (\ref{decomp}). For example 
the adjoint representation ${\bf 15}$ of SO(6) is decomposed as ${\bf 10}\oplus{\bf 5}$ under SO(5). Hence $\Psi$ in the representation ${\bf 15}$ is not the Killing scalar of the coset space SO(6)/SO(5)(=S$^5$). A similar comment can be done for the coset space SU($N$)/\{U(1)\}$^{N-1}$. The fundamental representation is decomposed into $N$ components each of which transforms under $\{U(1)\}^{N-1}$ having  U(1) charges  designated  by the weight vector $w_\mu,\ \mu=1,2,\cdots,w_N$. 
 Therefore $\Psi$ in the fundamental representation is not the Killing scalar. However it is worth noting that the elements of ${\cal D}(U)$ defined by (\ref{D(U)}) in the fundamental representation
 for SU($N$)/\{U(1)\}$^{N-1}$ were identical to the quantities  known as   the harmonic coordinates of the coset space discussed in \cite{West}. 
 The representation in which  $\Psi$ has the decomposition containing singlets  and becomes Killing scalar is the adjoint representation or the one obtained by its symmetric tensor product.  Indeed the adjoint representation contains $N-1$ null weight vectors. Note that the coset space SU($N$)/\{U(1)\}$^{N-1}$ is a K\"ahler manifold. $\Psi$ in the adoint representation  is nothing but the quantity as the Killing potential in the literature. The naming "Killing scalar" is based on this observation.

For the case where the representation $\mbox{\boldmath $N$}$ does not contain any singlet  we can 
 turn $\Psi$ given by (\ref{Psi}) to the Killing scalar in the following way.  Consider the Wilson line operator 
\begin{eqnarray}
W(\phi,\phi_0)=P\exp \int_{\phi_0}^\phi d\phi^a\omega_a^IH^I.   \nonumber
\end{eqnarray}
Here $\omega_a^IH^I$ is the connection defined from the Cartan-Maurer $1$-form as (\ref{Cartan}) and  transforms as
\begin{eqnarray}
\delta\Big(\omega^I_aH^I\Big)_{N_\nu\times N_\nu}= \Big(\partial_a\rho(\phi,\epsilon)-[\omega^I_aH^I,\rho(\phi,\epsilon)]\Big)_{N_\nu\times N_\nu} \nonumber
\end{eqnarray}
by (\ref{transf}). Hence  the Wilson line operator  transforms as 
$$
W(\phi,\phi_0) \longrightarrow e^{i\rho(\phi,\epsilon)}W(\phi,\phi_0)e^{-i\rho(\phi_0,\epsilon)}.
$$
by (\ref{transf}). 
The generator of the compensator $\rho(\phi,\epsilon)$ becomes $\epsilon^IH^I$ at the origin $\phi_0=0$ of the coset space G/H. 
Let $\eta$ to be  a linear representation vector  
$
e^{i\theta^I H^I}\eta_0
$
with $\theta^I$ parametrizing the subgroup H, so that it transforms by (\ref{transf})  as 
$$
 \eta \equiv e^{i\theta^I H^I}\eta_0
  \longrightarrow  
e^{i(\theta^I+\epsilon^I) H^I}\eta_0.
$$
Here $\eta_0$ is a constant vector which is fixed  in the representation space of $H^I$.  Then  we have  
\begin{eqnarray}
W(\phi,0)\eta \longrightarrow 
e^{i\rho(\phi,\epsilon)}W(\phi,0)\eta.   \nonumber
\end{eqnarray}
We have already known that  by (\ref{transf}) the quantity (\ref{Psi}) transforms  as $(\mbox{\boldmath $N$},\mbox{\boldmath $N$}^{w_\nu})$, i.e., (\ref{Psitransf}).  
As the result the following quantity  
\begin{eqnarray}
\Psi(x) W(\phi,0)\eta   \label{KillingSS}  
\end{eqnarray}
is the Killing scalar transforming as (\ref{KillingS}) in  any representation $\mbox{\boldmath $N$}$ of G. The point of the argument here is that the existence of a quantity 
 transforming as $(\bf 1,\mbox{\boldmath $N$}^{w_\nu})$ by (\ref{transf}) is not hypothetical, but it indeed exists as $W(\phi,0)\eta$.

It is here  opportune to discuss a relation between the vielebeine $e^i_a$ and the Killing vectors $R^{Aa}$ given by (\ref{Cartan}) and (\ref{Killing}) respectively. Let ${\cal D}(U)$ given by (\ref{D(U)}) to be the adjoint representation of G. As the adjoint representation is decomposed as (\ref{T}) we have
\begin{eqnarray}
{\cal D}(U)=\left(
\begin{array}{c|c}
(U)_{ij}  &  (U)_{iJ}   \\  
\vspace{-0.3cm}   \\  
\hline
\vspace{-0.3cm}   \\
(U)_{Ij}  &  (U)_{IJ}   
\end{array}
\right).  
   \label{DU}
\end{eqnarray}
It is an orthogonal matrix, so that the column vectors have length 1 and are orthogonal to the row vectors. Moreover from the transformation (\ref{transf}) we find  
\begin{eqnarray}
\left(
\begin{array}{c}
\hspace{-0.1cm} (U)^{ij} \hspace{-0.1cm}   \\
\vspace{-0.3cm}   \\
\hline
\vspace{-0.3cm}   \\
\hspace{-0.1cm} (U)^{Ij} \hspace{-0.1cm}     
\end{array}
\right)
 &\longrightarrow& 
\left(
\begin{array}{c|c}
(e^{\epsilon^AT^A})^{ik}  &  (e^{\epsilon^AT^A})^{iK}   \\
\vspace{-0.3cm}   \\
\hline
\vspace{-0.3cm}   \\
(e^{\epsilon^AT^A})^{Ik}  &  (e^{\epsilon^AT^A})^{IK}   
\end{array}
\right) 
\left(
\begin{array}{c}
\hspace{-0.1cm} (U)^{kl} \hspace{-0.1cm}   \\
\vspace{-0.3cm}   \\
\hline
\vspace{-0.3cm}   \\
\hspace{-0.1cm} (U)^{Kl} \hspace{-0.1cm}     
\end{array}
\right)
\Big( e^{-i\rho(\phi,\epsilon)}\Big)^{lj},  \nonumber  \\
\vspace{0.1cm}   \nonumber  \\
e^i_a d\phi^a &\longrightarrow& e^l_ad\phi^a 
\Big( e^{-i\rho(\phi,\epsilon)}\Big)^{li}.   \nonumber
\end{eqnarray} 
By using the Lie-variation we may write the second transformation   in the infinitesimal form 
\begin{eqnarray}
{\cal L}_{R^A}e^i_a \equiv R^{Ab}e^i_{a,b} +R^{Ab}_{\ \ ,a}e^i_b
= -\rho(\phi,\epsilon)^{ij} e^j_a .    \nonumber
\end{eqnarray}
Together with ${\cal L}_{R^A}R^{Bb}= f^{ABC}R^{Cb}$, given by (\ref{Lie-algebra}),  these observations lead us to the following relation between the vielebeine $e^i_a$ and the Killing vectors $R^{Aa}$
\begin{eqnarray}
R^{Aa}e^j_a\equiv 
\left(
\begin{array}{c}
\hspace{-0.1cm} R^{ia} e^j_a \hspace{-0.1cm}   \\
\vspace{-0.3cm}   \\
\hline
\vspace{-0.3cm}   \\
\hspace{-0.1cm} R^{Ia} e^j_a   \hspace{-0.1cm}     
\end{array}
\right)
=
\left(
\begin{array}{c}
\hspace{-0.1cm} (U)^{ij} \hspace{-0.1cm}   \\
\vspace{-0.3cm}   \\
\hline
\vspace{-0.3cm}   \\
\hspace{-0.1cm} (U)^{Ij} \hspace{-0.1cm}     
\end{array}
\right).    \label{Re} 
\end{eqnarray}
We note that
\begin{eqnarray}
 R^{Aa}|_{\phi=0}=\delta^{Aa}, \quad\quad e^i_a|_{\phi=0}=\delta^i_a,
\end{eqnarray}
by the construction. 

The metric of the coset space may be naively given by 
\begin{eqnarray}
 g^{ab}=R^{Aa}R^{Bb}, \quad\quad g_{ab}=e^i_ae^i_b.   \label{metric}
\end{eqnarray}
However this way of giving the metric may be generalized if the broken generators $X^i$ in (\ref{T}) are decomposed  as
$
\{X^{i_1},X^{i_2},\cdots, X^{i_n}\} 
$
under H, or equivalently the adjoint representation $\mbox{\boldmath $N$}^G$ is decomposed as
\begin{eqnarray}
\mbox{\boldmath $N$}^G= \mbox{\boldmath $N$}^{i_1}\oplus \mbox{\boldmath $N$}^{i_2}\oplus\cdots\oplus\mbox{\boldmath $N$}^{i_{n}}\oplus\mbox{\boldmath $N$}^H    \label{Cartan2} 
\end{eqnarray}
in the Cartan-Weyl basis. Here $\mbox{\boldmath $N$}^{i_r}, r=1,2,\cdots,n$, denote the $N_r$-dimensional representation of H, in which $i_r$ stands for a set of the roots for the representation. Hence (\ref{DU}) becomes 
\begin{eqnarray}
{\cal D}(U)=(U)_{N_G\times N_G}=
\left(
\begin{array}{ccc|c}
(U)_{\scriptscriptstyle{N_1\times N_1}}  & \cdots  & (U)_{\scriptscriptstyle{N_1\times N_n}}& (U)_{\scriptscriptstyle{N_1\times N_H}}   \\
\vspace{-0.3cm}  &  &  &    \\
 \hspace{-0.2cm}  \vdots   &  \ddots  &\hspace{-0.2cm} \vdots & \vdots  \\
\vspace{-0.3cm}  &  &  &    \\
(U)_{\scriptscriptstyle{N_n\times N_1}}  & \cdots  & (U)_{\scriptscriptstyle{N_n\times N_n}}& (U)_{\scriptscriptstyle{N_n\times N_H}}   \\
 \vspace{-0.3cm}     &   &    &    \\
\hline
\vspace{-0.3cm}  &  &  &    \\
(U)_{\scriptscriptstyle{N_H\times N_1}}  & \cdots  & (U)_{\scriptscriptstyle{N_H\times N_n}}& (U)_{\scriptscriptstyle{N_H\times N_H}}   
\end{array}
\right),    \label{DUU}
\end{eqnarray}
with $N^{\rm G}=dim$ G and $N^{\rm H}=dim$ H. Then the metric (\ref{metric}) can  be generalized to 
\begin{eqnarray}
g^{ab}=R^{Aa}(UPU^{-1})^{AB}R^{Bb}, \quad\quad\quad 
g_{ab}=\eta^{ij}e^i_ae^j_b,  \label{metricgene}
\end{eqnarray}
with a projection  operator $P$ such as 
\begin{eqnarray}
P =
\left(
\begin{array}{c|c}
(P)_{ij}  &  (P)_{iJ}   \\  
\vspace{-0.3cm}   \\  
\hline
\vspace{-0.3cm}   \\
(P)_{Ij}  &  (P)_{IJ}   
\end{array}
\right)  
=  
\left(
\begin{array}{c|c}
(\eta^{-1})_{ij}  & \hspace{0.5cm} 0 \hspace{0.5cm}  \\  
\vspace{-0.3cm}   \\  
\hline
\vspace{-0.3cm}   \\
 0  &  \hspace{0.5cm} 0 \hspace{0.5cm} 
\end{array}
\right).   \nonumber
\end{eqnarray}
Here $\eta^{ij}$ is given by 

\newfont{\bg}{cmr10 scaled\magstep4}
\newcommand{\bigzerol}{\smash{\hbox{\bg 0}}}
\newcommand{\bigzerou}{%
 \smash{\lower1.7ex\hbox{\bg 0}}}

\begin{eqnarray}
\eta^{ij} = 
\left(
\begin{array}{ccc}
\hspace{-0.1cm} ({ c_1})_{N_1\times N_1} & \hspace{-0.1cm}\cdots &\hspace{-0.2cm}  \bigzerou \hspace{-0.1cm} \\
\noalign{\vskip0.2cm} 
\hspace{-0.1cm} \vdots   &\hspace{-0.1cm}\ddots    &\hspace{-0.2cm} \vdots \hspace{-0.1cm}   \\
\noalign{\vskip0.3cm}
\hspace{-0.1cm} \bigzerol     & \hspace{-0.1cm}\cdots     &   ({ c_n})_{N_n\times N_n}  \hspace{-0.1cm} \\
\end{array}
\right),   \nonumber 
\end{eqnarray}
in which we have $(c_r)_{N_r\times N_r}=c_r({\bf 1})_{N_r\times N_r}$ with some constants $c_r, r=1,\cdots,n$\cite{Ao4}. 
It is worth noting that the inverse of the vielebeine $e^i_a$ are given by 
$$
e^{ai}=R^{Aa}U^{Ai}.
$$

\section{The classical exchange algebra}
\setcounter{equation}{0}

We are now in a position to discuss the classical exchange algebra of the ordinary non-linear $\sigma$-model on G/H. 
 The non-linear $\sigma$-model on G/H is given by 
\begin{eqnarray}
S=\int d^2\xi\ {\cal L}={1\over 2}\int d^2\xi\ \eta^{\mu\nu}g_{ab}(\phi)\partial_\mu \phi^a\partial_\nu \phi^b,
  \nonumber
\end{eqnarray}
in  the two-dimensional flat world-sheet.  Here use is made of the formula (\ref{metric}) or (\ref{metricgene}) for the metric $g_{ab}(\phi)$. 
We study the Poisson structure on the light-like plane $x^+=y^+$.  
The Dirac method hardly works to set up the Poisson brackets $\displaystyle{\{\phi^a(x)\mathop{,}^\otimes \phi^b(y)\}}$, because there appear the first  and send class constraints, which we do not know how to disentangle. The reader refer to \cite{Ao3} for a detailed argument on this. Hence we assume that the Poisson brackets can be set up so as to satisfy the two requirements. In the first place  the energy-momentum tensor $T_{--}$ should reproduce  diffeomorphism as 
\begin{eqnarray}
\delta_{diff} \phi^a(x^+,x^-) &\equiv&\epsilon(x^-)\partial_- \phi^a(x^+,x^-)  \nonumber\\
          &=& \int dy^- \epsilon(y^-)\{\phi^a(x), T_{--}(\phi(x))\}\Big |_{x^+=y^+},
 \label{diffeo}    
\end{eqnarray}
by using the Poisson brackets $\displaystyle{\{\phi^a(x)\mathop{,}^\otimes \phi^b(y)\}}$. Secondly they should be consistent with the Jacobi identities. 
These  requirements are  satisfied with the Poisson brackets of the following form 
\begin{eqnarray}
&\ & \{\phi^a(x)\mathop{,}^\otimes \phi^b(y)\}   \nonumber \\
&\ &\quad\quad =-{1\over 4 }\Big[\theta(x-y)t_{AB}^+ \delta^A\phi^a(x)\otimes \delta^B\phi^b(y) -\theta(y-x)t_{AB}^+ \delta^A\phi^b(y)\otimes \delta^B\phi^a(x)\Big]\quad\     \label{Poisson}
\end{eqnarray}
on the light-like plane $x^+=y^+$. 
The notation is as follows. $\theta(x)$ is the step function. $\delta^A\phi^a(x)$ are the Killing vectors defined by (\ref{Killing}). More correctly they should be written as $\delta^A\phi^a((\phi(x))$, but the dependence of $\phi(x)$ was omitted to avoid an unnecessary complication. The quantity $t^+_{AB}$ is  the most crucial in our arguments. It is a modified Killing metric which defines the classical r-matrices as 
\begin{eqnarray}
 r^{\pm}=\sum_{\alpha\in R} sgn\ \alpha E_{\alpha}\otimes E_{-\alpha} \pm
   \sum_{A,B} t_{AB}T^A \otimes T^B   
    \equiv  t^\pm_{\ AB}T^A\otimes T^B, \label{r}
\end{eqnarray}
with  $T^A$ the generators of the group G given in the Cartan-Weyl basis as $\{E_{\pm\alpha},H_\mu\}$, $t_{AB}$ the corresponding Killing metric and 
$sgn\ \alpha=\pm$ according as the roots are positive or negative. Note the relation $t_{AB}^+=-t_{BA}^-$. The r-matrix satisfies the classical Yang-Baxter equation (\ref{YB})\cite{J}. 
It is easy to show that the first requirement (\ref{diffeo}) is satisfied with the Poisson brackets (\ref{Poisson}). The second requirement can be similarly shown as has been done  in \cite{Ao1,Ao3}. (See eqs. (3.20) and (12) in the respective reference.) 

Note that 
$$
\{\phi^a(x)\mathop{,}^\otimes \Psi(y)\}=\{\phi^a(x)\mathop{,}^\otimes \phi^b(y)\}{\delta \Psi(y)\over \delta \phi^b(y)}.
$$ 
By using the Poisson brackets (\ref{Poisson}) 
we can easily show the Killing scalar defined   by (\ref{Psi}) to satisfy 
 the classical exchange algebra in the form
\begin{eqnarray}
\{\Psi(x)\mathop{,}^\otimes \Psi(y)\}=-{1\over 4}[\theta(x-y)r^+ + \theta(y-x) r^-]\Psi(x)\otimes\Psi(y),   \label{CEA}
\end{eqnarray}
on the light-like plane $x^+=y^+$. 
Here $\Psi(x)$ should be understood with an abbreviated notation for $\Psi(\phi(x))$. It is also understood as generalized by means of (\ref{KillingSS}), when  its representation of G does not contain any singlet in the decomposition under H.

\section{Conclusions}

Finally we would like to comment on the S-matrix for the PSU(2$|$2) spin-chain system. found by Beisert\cite{Bei2}. The S-matrix is related to the R-matrix as $S=P\circ R$ with the permutation map $P$. 
Let us multiply $P$ on the quantum exchange algebra (\ref{QEA}) and the Yang-Baxter equation (\ref{YBB}). Calculate the l.h.s. as $PR_{yx}P^{-1}P\Psi(y)\otimes \Psi(x)$. Then the respective formula becomes 
\begin{eqnarray}
S_{xy}\Psi(x)\otimes \Psi(y)=\Psi(y)\otimes \Psi(x), 
\quad\quad\quad S_{yx}S_{zx}S_{zy}=S_{zy}S_{zx}S_{yx}.
  \nonumber
\end{eqnarray} 
In \cite{Bei2} the S-matrix was found by numerically solving the Yang-Baxter equation. 
Actually the solution had the extended symmetry PSU(2$|$2)$\ltimes\mathbb{R}^3$.  The classical r-matrix was discussed  from this S-matrix with an appropriate deformation parameter in \cite{Tori2}. The resulting r-matrix is not the kind which follows from  the Poisson structure of some underlying theory for the spin-chain system.

In this letter we have studied the Poisson structure of a generic non-linear $\sigma$-model on G/H formulating  on the light-like plane. Setting up the fundamental Poisson brackets (\ref{Poisson}) we have found the Killing scalar satisfying the classical exchange algebra (\ref{CEA0}) in an arbitrary representation, i.e., (\ref{CEA}). It is natural to think of 
a certain non-linear $\sigma$-model with the PSU($2|2$) or PSU($2,2|4$) symmetry  as a underlying world-sheet theory for the spin-chain system\cite{Bei1,Bei2}.  We need a special care in order to generalize our arguments to the case where 
the symmetry admits  non-trivial extension as PSU(2$|$2)$\ltimes\mathbb{R}^3$.  The exchange algebra for such a  non-linear $\sigma$-model   will be discussed in a future publication.

\end{document}